\documentclass[11pt]{article}
\usepackage{moriond,epsfig}

\usepackage[super,sort&compress]{natbib}
\usepackage{amsmath,amssymb}
\usepackage{graphicx}

\newcommand{\ev}[1]{\ensuremath{\left\langle #1 \right\rangle}} 

\def\MyPhoto{\pdfimage width 35mm {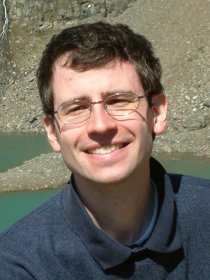}}

\def\be{\begin{equation}}
\def\ee{\end{equation}}
\def\bea{\begin{eqnarray}}
\def\eea{\end{eqnarray}}

\allowdisplaybreaks

\begin{document}
\vspace*{4cm}
\title{COLLIDER LIMITS ON DARK MATTER}

\author{J. KOPP \footnote{Based on work done in collaboration with Yang Bai,
        Patrick Fox, Roni Harnik and Yuhsin Tsai}}

\address{Theoretical Physics Department, Fermilab, PO Box 500, Batavia, IL 60510, USA}

\maketitle\abstracts{
Dark matter pair production at high energy colliders may leave observable
signatures in the energy and momentum spectra of the objects recoiling against
the dark matter. We discuss signatures of Dark Matter in the jets + missing
energy and photon + missing energy channels at the Tevatron and at
LEP. Working in a largely model-independent effective theory framework, we can
convert the collider bounds into constraints on the dark matter--nucleon
scattering cross section and on the dark matter annihilation cross section. Our
bounds are highly competitive with those from direct and indirect dark matter
searches, especially for light WIMPs and for WIMPs with spin-dependent or
leptophilic interactions. For example, we show that LEP rules out light
($\lesssim 10$~GeV) thermal relic dark matter if annihilation into electrons
is among the dominant annihilation channels.}

\section{Introduction}
\label{sec:intro}

Collider searches for dark matter are highly complementary to direct searches
looking for dark matter--nucleon scattering and to indirect searches looking
for signatures of dark matter annihilation or decay in stars or galaxies. The
main advantage of collider searches is that they do not suffer from
astrophysical uncertainties and that there is no lower limit to the dark matter
masses to which they are sensitive.

%

In this talk, we discuss search strategies for dark matter at colliders and
compare the obtained limits to those from direct and indirect searches. We work
in a largely model-independent effective field theory framework, assuming the
interactions between a dark matter Dirac fermion $\chi$ and standard model
fermions $f$ to be well described by contact operators of the form
\begin{align}
  \mathcal{O}_V &= \frac{(\bar\chi\gamma_\mu\chi)(\bar f \gamma^\mu f)}{\Lambda^2} \,,
    & \text{(vector, $s$-channel)} \label{O1} \\
  \mathcal{O}_S &= \frac{(\bar\chi\chi)(\bar f f)}{\Lambda^2} \,,
    & \text{(scalar, $s$-channel)} \\
  \mathcal{O}_A &= \frac{(\bar\chi\gamma_\mu\gamma_5\chi)(\bar f \gamma^\mu\gamma_5 f)}{\Lambda^2} \,,
    & \text{(axial vector, $s$-channel)} \label{O2} \\
  \mathcal{O}_t &= \frac{(\bar\chi f)(\bar f \chi)}{\Lambda^2} \,.
    & \text{(scalar, $t$-channel)} \label{O3}
\end{align}
While this set of operators is not exhaustive, it encompasses the essential
phenomenologically distinct scenarios: spin dependent and spin independent dark
matter--nucleus scattering, as well as $s$- and $p$-wave annihilation. The
classification of the effective operators as $s$-channel or $t$-channel refers
to the renormalizable model from which they typically arise:
\eqref{O1}--\eqref{O2} are most straightforwardly obtained if dark matter pair
production is mediated by a new neutral particle propagating in the
$s$-channel, while eq.~\eqref{O3} arises naturally if the mediator is a charged
scalar exchanged in the $t$-channel (for instance a squark or slepton). With
such a UV completion in mind, the suppression scale $\Lambda$ can be
interpreted as the mass of the mediator $M$, divided by the geometric mean of
its couplings to standard model fermions, $g_f$, and dark matter, $g_\chi$:
$\Lambda = M / \sqrt{g_f g_\chi}$. Note that there is some degree of redundancy
in eqs.~\eqref{O1}--\eqref{O3} because $\mathcal{O}_t$ can be rewritten as a
linear combination of $s$-channel type operator using the Fierz identities.

\begin{figure}
  \begin{center}
    \begin{tabular}{c@{\qquad\qquad}c@{\qquad\qquad}c}
      \includegraphics[width=2.7cm]{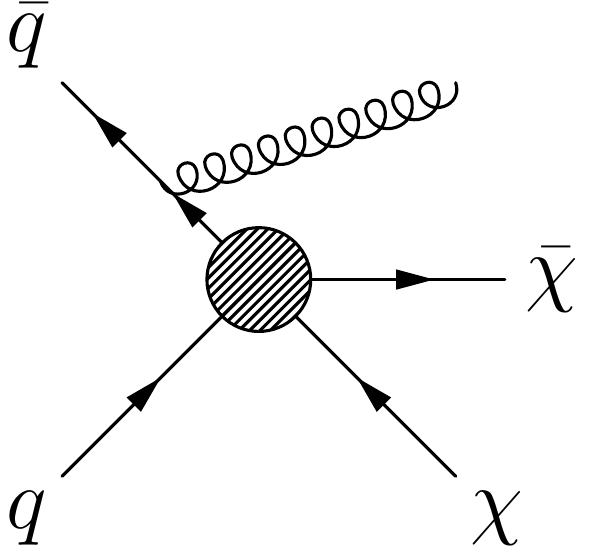} &
      \includegraphics[width=2.7cm]{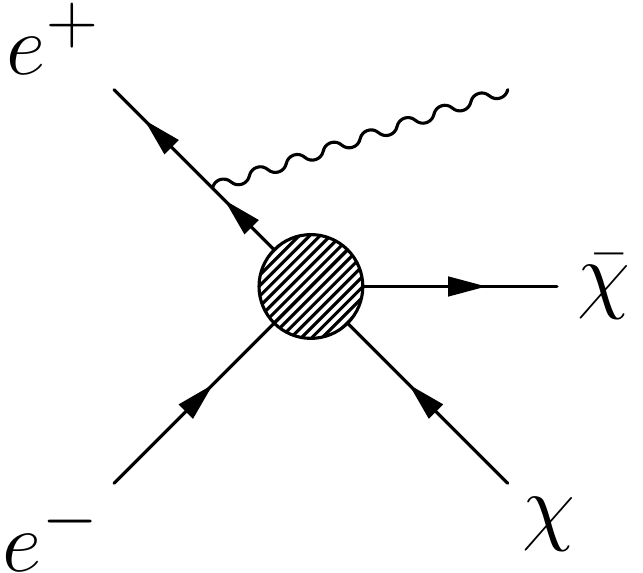} &
      \raisebox{-0.3cm}{\includegraphics[width=4.5cm]{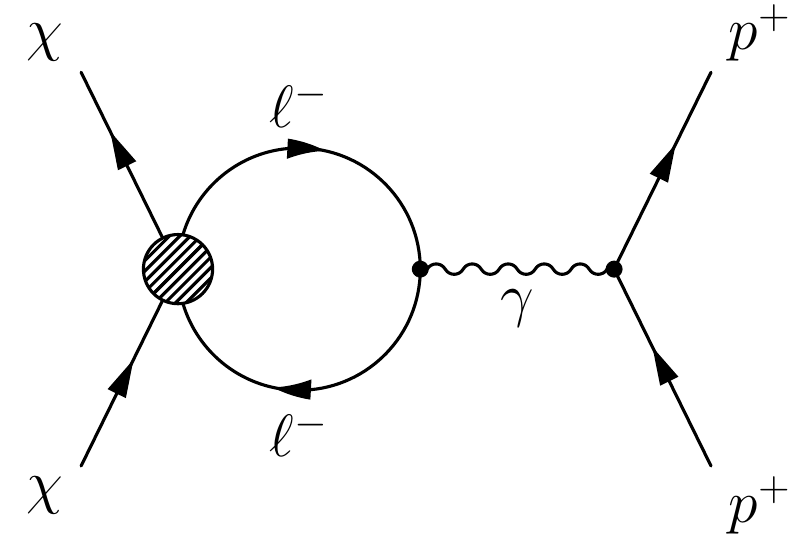}} \\
      (a) & (b) & (c)
    \end{tabular}
  \end{center}
  \caption{Dark matter production in association with (a) a mono-jet at a hadron
    colliders or with (b) a mono-photon at LEP. (c) Dark matter--nucleon scattering
    at one loop in models of leptophilic dark matter.}
  \label{fig:feyn}
\end{figure}

The experimental signatures we will investigate include events with a single
jet or a single photon and a large amount of missing energy
(fig.~\ref{fig:feyn} (a) and (b)).  In sec.~\ref{sec:tevatron}, we will focus
on searches at the Tevatron~\cite{Bai:2010hh,Goodman:2010ku,Goodman:2010yf},
while in sec.~\ref{sec:lep}, we will derive limits from a reanalysis of LEP
data~\cite{Fox:2011fx}.

\section{Mono-jets at the Tevatron}
\label{sec:tevatron}

Events in which dark matter is pair-produced can contribute to mono-jet events
at CDF\cite{Aaltonen:2008hh} through diagrams like the one in
fig.~\ref{fig:feyn} (a). By comparing the number of observed mono-jet events to
the number of events expected from dark matter production and from standard
model backgrounds, one can derive limits on the suppression scale $\Lambda$ of
the effective dark matter couplings as a function of the dark matter mass
$m_\chi$. These limits can then be converted into constraints on the dark
matter--nucleon scattering cross section.

\begin{figure}
  \begin{center}
    \includegraphics[width=7.5cm]{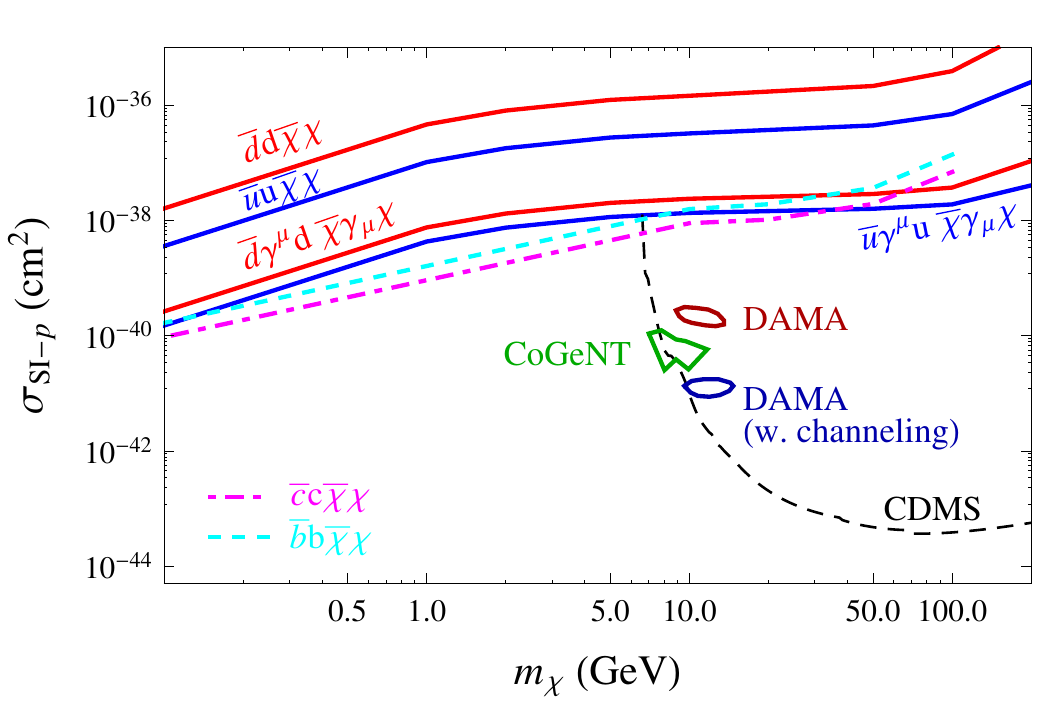} \hfill
    \includegraphics[width=7.5cm]{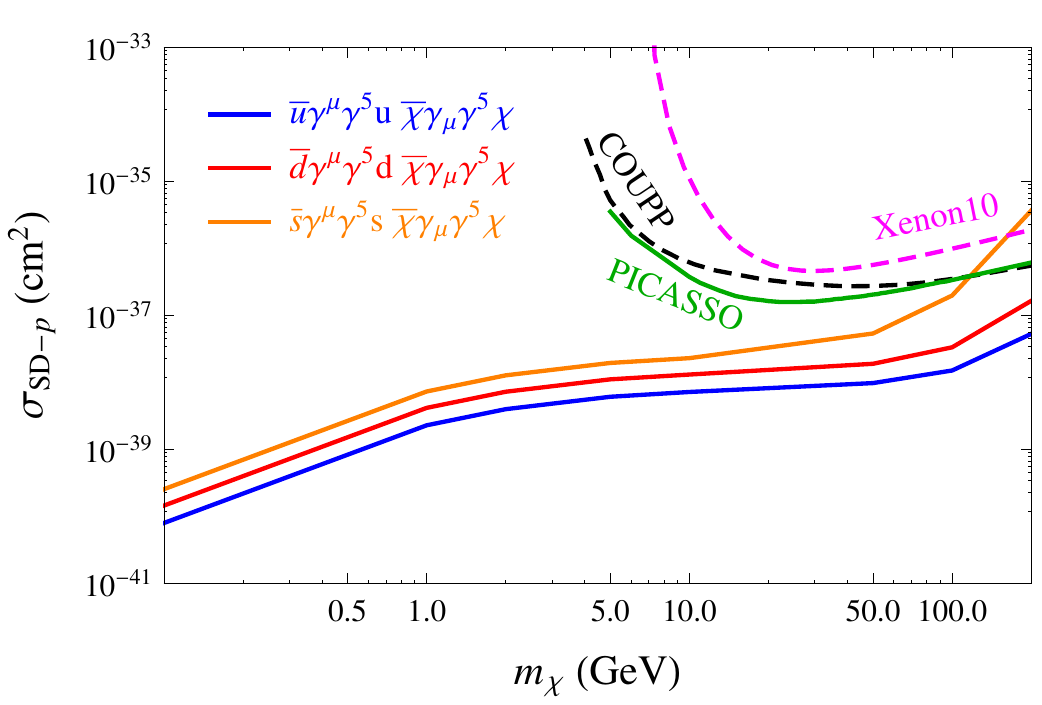}
  \end{center}
  \vspace{-0.2cm}
  \caption{Limits on spin-independent (left) and spin-dependent (right) dark
    matter--proton interactions from a Tevatron mono-jet
    search~\cite{Aaltonen:2008hh, Bai:2010hh}. We also show constraints from
    direct searches~\cite{Bernabei:2008yi, Ahmed:2009zw,
    Aalseth:2010vx}. Plots taken from Bai et al.~\cite{Bai:2010hh}.}
  \label{fig:dd-tev}
\end{figure}

In fig.~\ref{fig:dd-tev}, we compare these constraints to the ones obtained
from direct dark matter searches. We find that the Tevatron limits are stronger
than those from direct searches if dark matter is lighter than a few GeV or has
predominantly spin-dependent interactions.  At $m_\chi \sim \text{few} \times
100$~GeV, the Tevatron's sensitivity deteriorates due to kinematic limitations.

Note that the Tevatron mono-jet search is limited by systematic uncertainties,
so more data alone will not be sufficient to improve the limits considerably.
However, some improvement can be expected from an analysis taking into account
not only the total number of mono-jet events, but also the transverse momentum
spectrum of the jets. Such an analysis would require good understanding of the
uncertainties associated with the prediction of QCD backgrounds.  Performing an
inclusive rather than exclusive search may help to reduce the these
uncertainties.


\section{Mono-photons at LEP}
\label{sec:lep}

\begin{figure}
  \begin{center}
    \includegraphics[width=7.5cm]{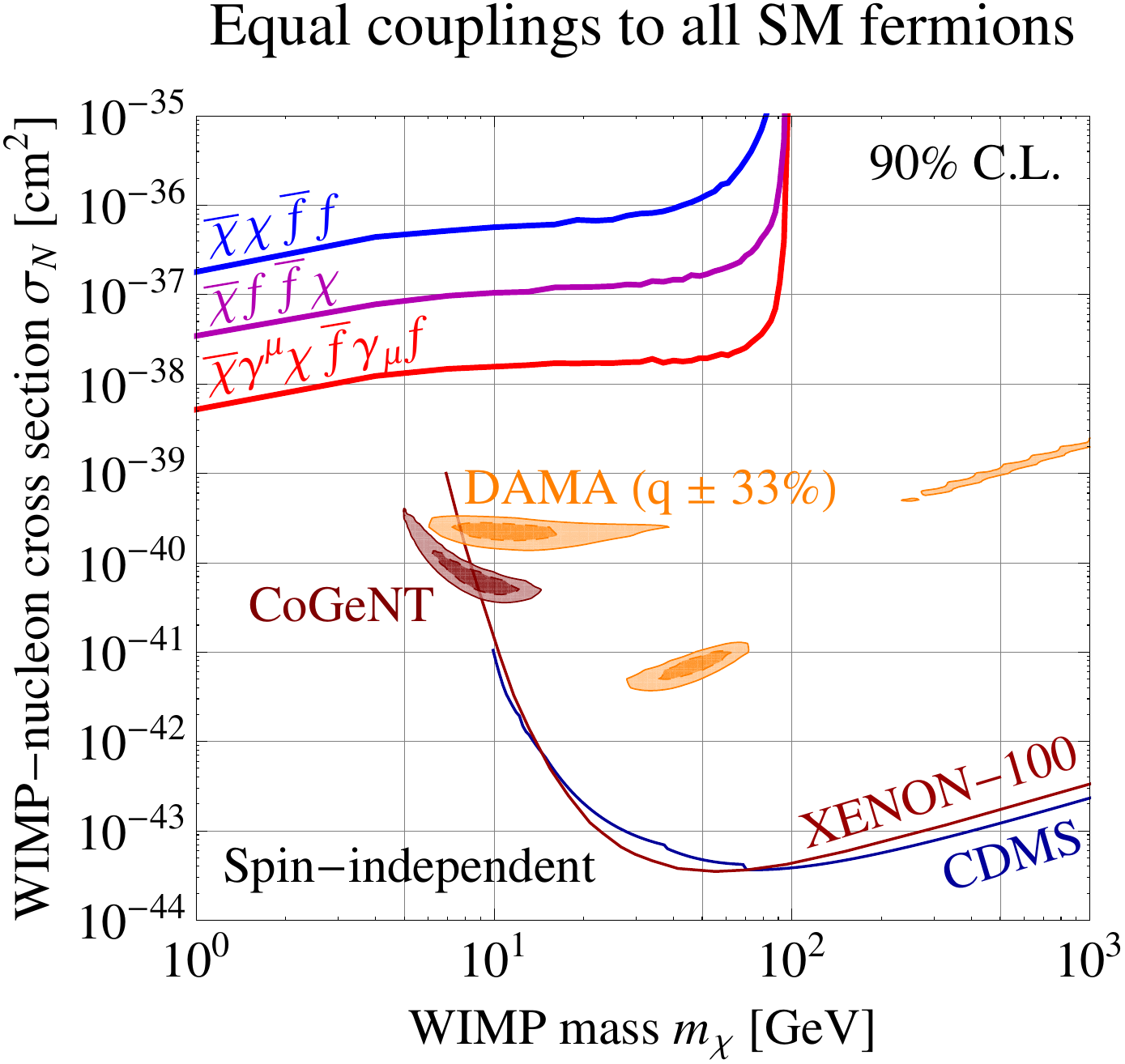} \hfill
    \includegraphics[width=7.5cm]{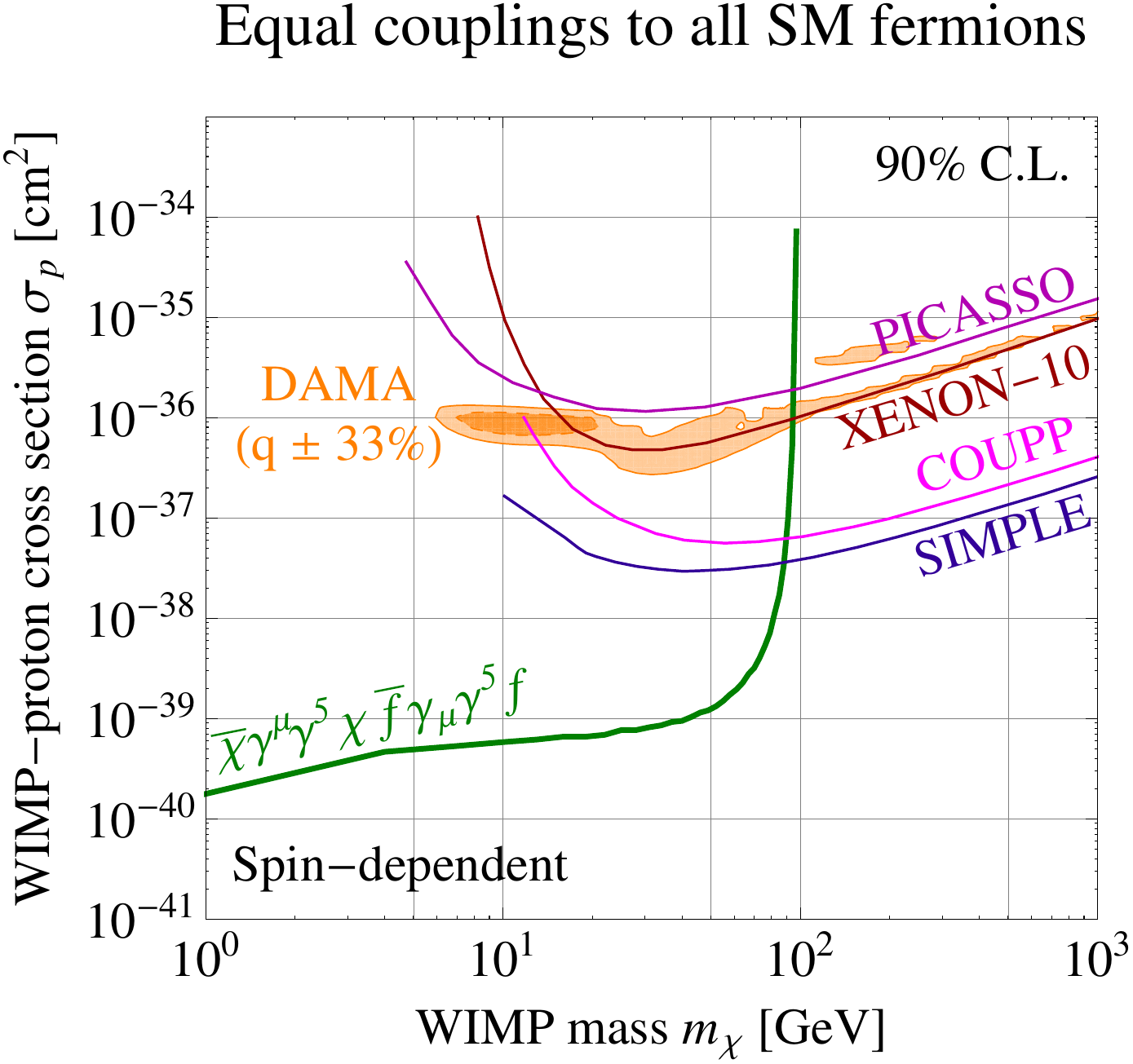}
  \end{center}
  \vspace{-0.2cm}
  \caption{Limits on spin-independent (left) and spin-dependent (right) dark
    matter--nucleon interactions from a LEP mono-photon
    search~\cite{DELPHI:2008zg, Fox:2011fx}. Results are compared to
    constraints from various direct searches~\cite{Bernabei:2008yi,
    Ahmed:2009zw, Aalseth:2010vx, Aprile:2010um, BarnabeHeider:2005pg,
    Angle:2008we, Behnke:2010xt, Girard:2011xc, Kopp:2009qt}.}
  \label{fig:dd-lep}
\end{figure}

Even though the total integrated luminosity of around $650$~pb$^{-1}$ recorded
by the LEP experiments is smaller than the data set available at the Tevatron,
we will now show that this data can still be used to set highly competitive
limits on the properties of dark matter. Since initial state QCD radiation is
absent at LEP, we will focus on final states with a single photon and a large
amount of missing energy, {\it i.e.}\ we will study the process $e^+ e^- \to
\bar\chi \chi \gamma$ (fig.~\ref{fig:feyn} (b)). Our analysis is based on the
mono-photon spectrum observed by the DELPHI detector~\cite{Abdallah:2003np,
DELPHI:2008zg}, which we will compare to predictions obtained using
CompHEP~\cite{Boos:2004kh}, together with our own implementation of the DELPHI
cuts, efficiencies, and energy resolutions in a modified version of the
MadAnalysis framework~\cite{Alwall:2007st}. Details on technical aspects of our
analysis can be found in ref.~\cite{Fox:2008kb}.  We have verified our
simulations by checking that we are able to reproduce the mono-photon
distribution expected from the background process $e^+ e^- \to Z \gamma$, with
the $Z$ decaying invisibly.


Like for the mono-jet channel, we first derive limits on the suppression scale
$\Lambda$ as a function of $m_\chi$. While for mono-jets, a spectral analysis
would have required detailed understanding of the systematic uncertainties in
the background prediction, the mono-photon search at LEP is statistics-limited,
and it is therefore straightforward to take into account the full mono-photon
spectrum. This is advantageous because the distribution of signal events
expected from dark matter pair production is different from the shape of the
$e^+ e^- \to \bar\nu \nu \gamma$ background.

To convert the LEP bounds on $\Lambda$ into limits on the dark matter--nucleon
scattering cross section, we need to make some assumption on the relative
strength of dark matter--quark couplings compared to dark matter--electron
couplings. If these couplings are identical, as assumed in
fig.~\ref{fig:dd-lep}, we find that the collider limits are again highly
competitive for very light dark matter ($m_\chi \lesssim 4$~GeV) and for
spin-dependent scattering up to the kinematic cut-off of LEP.

\begin{figure}
  \begin{center}
    \includegraphics[width=7.5cm]{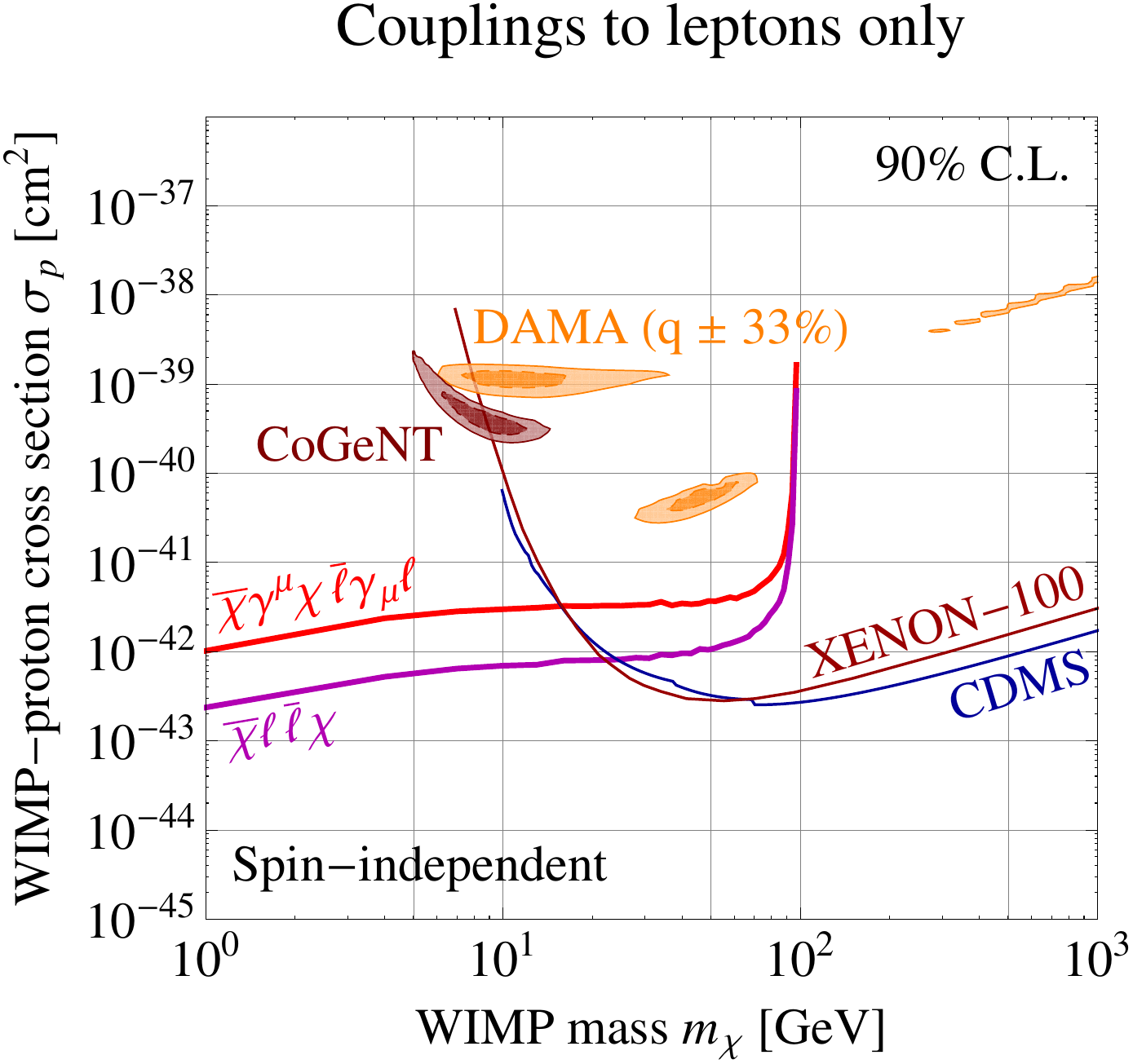}
  \end{center}
  \vspace{-0.2cm}
  \caption{LEP constraints on dark matter with tree level couplings
    only to leptons~\cite{Fox:2011fx}, compared to limits from direct detection
    experiments~\cite{Bernabei:2008yi, Ahmed:2009zw, Aalseth:2010vx,
    Aprile:2010um, Kopp:2009et, Kopp:2010su}.}
  \label{fig:leptophilic}
\end{figure}

LEP can do even better in models where dark matter is leptophilic, {\it i.e.}\ has
tree level couplings predominantly to leptons. Such models are, for example,
motivated by recent anomalies in cosmic ray spectra.\cite{Adriani:2008zr,
Abdo:2009zk}  Even though dark matter--nucleon scattering may be absent or
suppressed in such models at the tree level, it can still occur at the loop
level, mediated for instance by the diagram shown in fig.~\ref{fig:feyn}
(c).\cite{Kopp:2010su} The expected signal in direct detection experiments in
this case is suppressed by a loop factor, so that LEP, which is probing
unsuppressed tree level interactions, has a relative advantage and is
competitive with direct searches even for spin-independent scattering up to its
kinematic limit around $m_\chi \sim 80$~GeV (see fig.~\ref{fig:leptophilic}).

\begin{figure}
  \begin{center}
    \begin{tabular}{cc}
      \includegraphics[width=7.5cm]{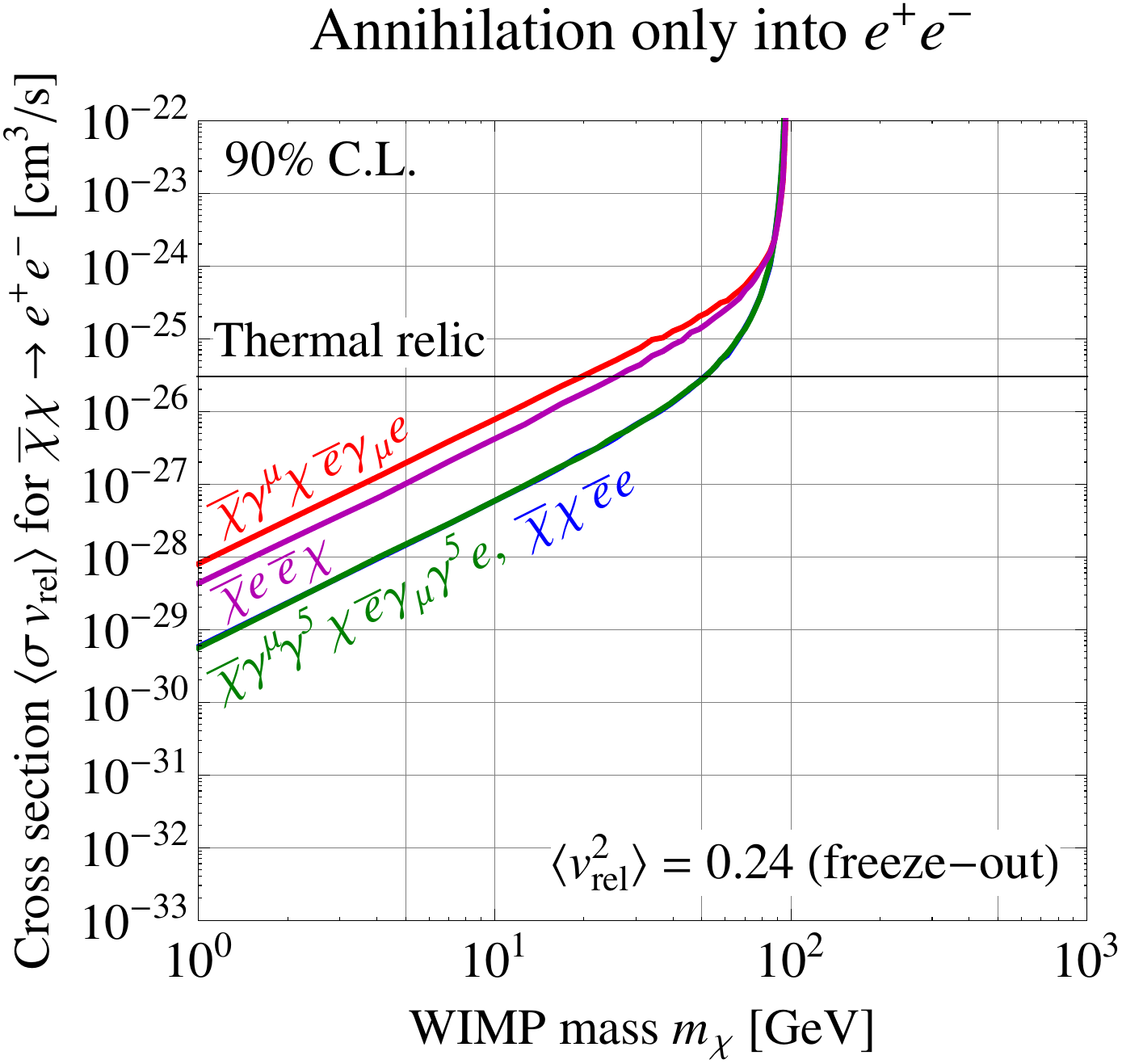} &
      \includegraphics[width=7.5cm]{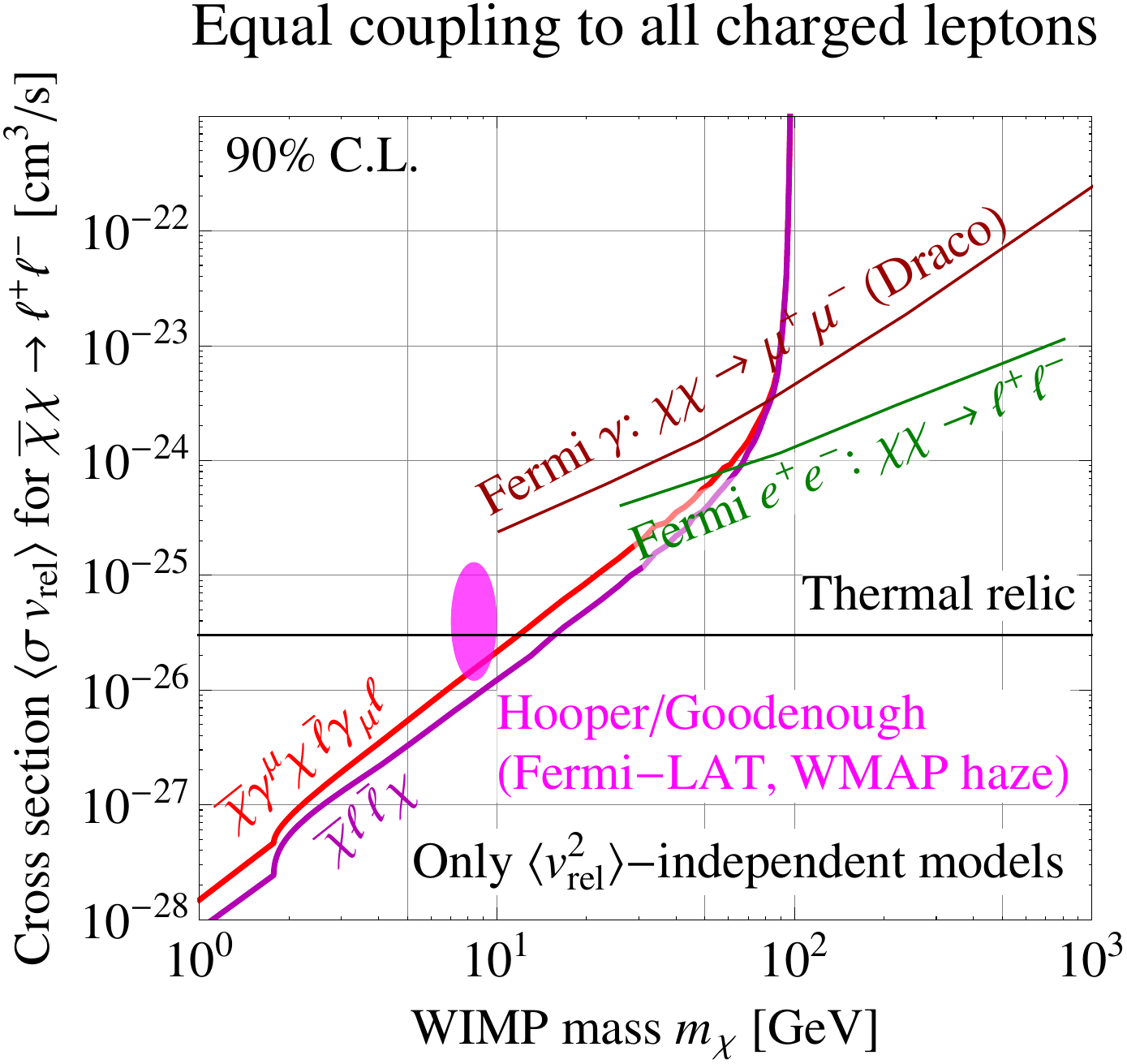} \\
      (a) & (b)
    \end{tabular}
  \end{center}
  \vspace{-0.2cm}
  \caption{LEP constraints on the dark matter annihilation cross section
    for the case where the branching ratio for $\bar\chi \chi \to e^+ e^-$
    is 100\% (left), and for the case where dark matter couples equally to all
    charged leptons (right)~\cite{Fox:2011fx}.}
  \label{fig:annihilation}
\end{figure}

Besides the dark matter--nucleon scattering cross section, LEP can also set
limits on the dark matter annihilation cross section. {\it Per se}, only bounds
on annihilation into $e^+ e^-$ pairs can be derived
(fig.~\ref{fig:annihilation} (a)), but it is easy to generalize these bounds,
though not in a model-independent way. In particular, if there are other
annihilation channels besides $\bar\chi \chi \to e^+ e^-$, the LEP limits on
the annihilation cross section are weakened by the inverse of the branching
ratio for $\bar\chi \chi \to e^+ e^-$.  Since the cross sections for some types
of dark matter interactions (in particular scalar and axial vector) depend
strongly on the relative velocity $v_{\rm rel}$ of the annihilating dark matter
particles, we have to specify the value of this quantity.  In
fig.~\ref{fig:annihilation}, we take the average squared velocity $\ev{v_{\rm
rel}^2}$ to have a value of 0.24, corresponding to the time of electron--proton
recombination in the early universe.  (At later times, $\ev{v_{\rm rel}^2}$ is
smaller and the limits on scalar and axial vector interactions improve
dramatically.\cite{Fox:2011fx}) We see that, if dark matter annihilates
exclusively into $e^+ e^-$ pairs, LEP is able to rule out the annihilation
cross section required for thermal relic dark matter, $\ev{\sigma v_{\rm rel}}
= 3 \times 10^{-26}$~cm$^3$/s, if $m_\chi \lesssim \mathcal{O}(10\
\text{GeV})$.

In fig.~\ref{fig:annihilation} (b) we compare LEP limits on the dark matter
annihilation cross section to various astrophysical
constraints~\cite{Abdo:2010ex, Grasso:2009ma, Hooper:2010mq}. We assume dark
matter to couple equally to all charged leptons, but it would again be
straightforward to rescale our limits if this is not the case. We see that for
low $m_\chi$ LEP limits are stronger than constraints from gamma ray and $e^+
e^-$ observations by the Fermi-LAT collaboration, and that LEP is also able to
disfavor a large portion of the parameter region that could potentially explain
gamma ray signals from the galactic center.\cite{Hooper:2010mq}

\begin{figure}
  \begin{center}
    \includegraphics[width=7.5cm]{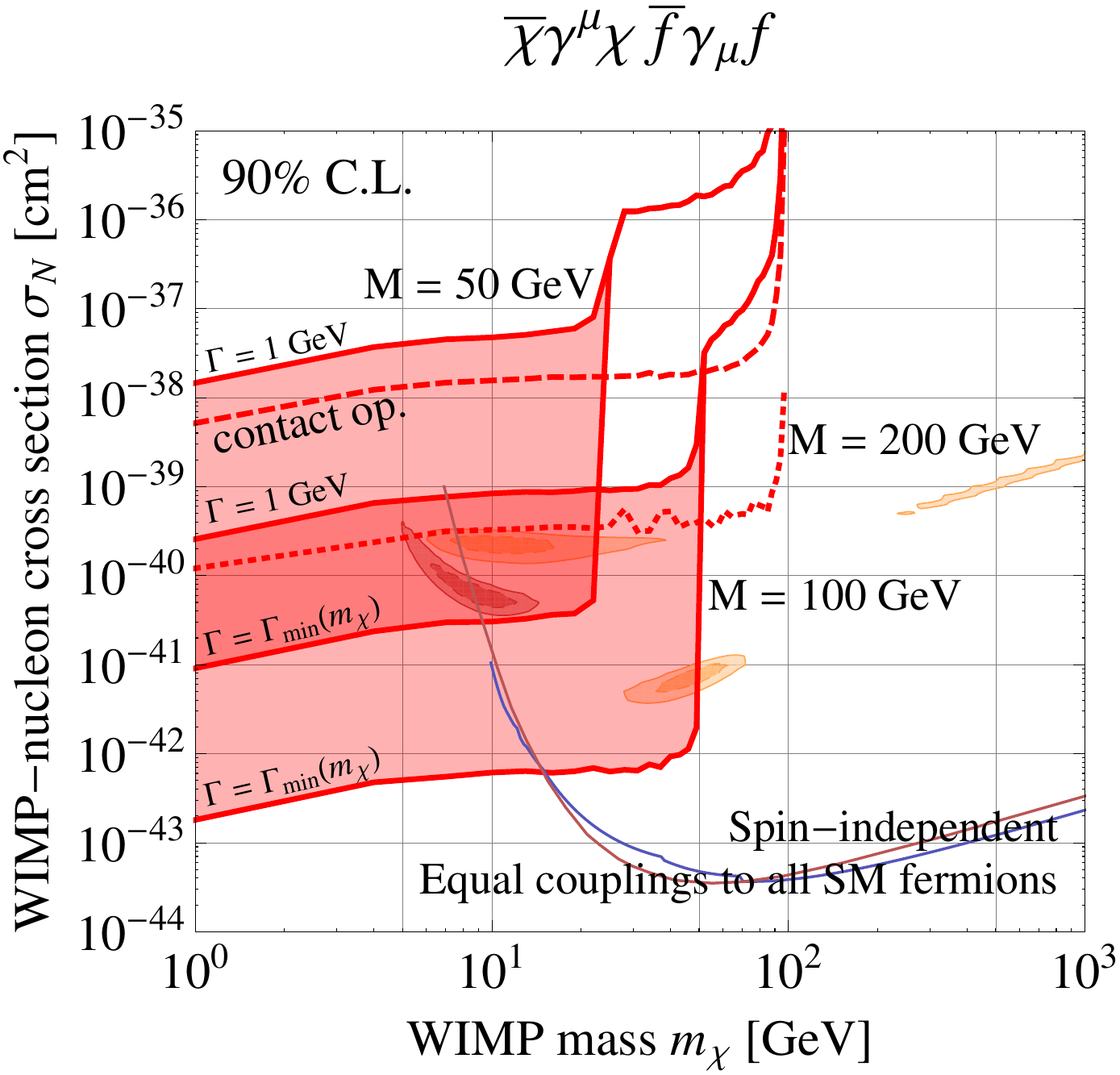}
  \end{center}
  \vspace{-0.2cm}
  \caption{LEP constraints on the dark matter--nucleon scattering cross section
    in models where the interactions are mediated by a relatively light
    particle~\cite{Fox:2011fx}.}
  \label{fig:lightM}
\end{figure}

In fig.~\ref{fig:lightM}, we depart from the effective theory formalism and
consider the implications of dark matter interactions mediated by a particle
whose mass $M$ is comparable to or below the LEP center of mass energy
$\sqrt{s} \sim 200$~GeV. For $M \sim \sqrt{s}$, there is a regime where dark
matter production at LEP is resonantly enhanced, so that the limit on the dark
matter--nucleon scattering cross section $\sigma_N$ improves compared to the
contact operator case.  For smaller $M$, the LEP constraint becomes generally
weaker because the production cross section at LEP is proportional to $s^{-1}$,
whereas $\sigma_N$ is proportional to  $\mu_N^2 / M^4$ (with the dark
matter--nucleon invariant mass $\mu_N$), giving direct detection experiments a
relative advantage at small $M$. A special situation arises when $2 m_\chi <
M$, so that the mediator can be produced on-shell at LEP and then decay into
dark matter. In that case, the LEP limit on $\sigma_N$ is very sensitive to the
width $\Gamma$ of the mediator, which is a measure for its branching ratio into
$\bar\chi \chi$ (larger $\Gamma$ implies smaller branching ratio). We also note
that on-shell production of the mediator with subsequent decay into standard
model particles may impose independent constraints on models of this type.

\section{Conclusions}
\label{sec:conclusions}

In conclusion, we have shown that a largely model-independent search for dark
matter is possible at high-energy hadron and lepton colliders by looking for an
excess of events with large missing energy and a single jet or photon from
initial state radiation.  Working in an effective field theory framework, we
have shown that the limits that LEP and the Tevatron can set on the mass and
couplings of dark matter are superior to direct detection constraints if dark
matter is very light ($\lesssim 4$~GeV) or has predominantly spin-dependent or
leptophilic interactions.  Above masses of $\mathcal{O}(100\ \text{GeV})$,
collider limits deteriorate due to kinematic limitations.  We have also used
LEP data to set limits on the dark matter annihilation cross section. For
example, we were able to rule out a thermal relic with a mass below $\lesssim
10$~GeV if the $e^+ e^-$ final state is among the dominant annihilation
channels.  Our limits on dark matter annihilation are highly complementary to
those from astrophysical searches since they extend to very low dark matter
masses, whereas astrophysical experiment are most sensitive for dark matter
masses above $\sim 50$~GeV.  Finally, we have also considered models in which
dark matter interactions are mediated by a light particle and thus cannot be
described in effective field theory. In this case, collider
constraints can weaken, but depending on the details of the model may also
become much stronger.


\bibliographystyle{kpmoriond}
\bibliography{collider-dm}

%
%
%
%

\end{document}